# An experimental system for detection and localization of hemorrhage using ultra-wideband microwaves with deep learning

Eisa Hedayati[1,2]*, Fatemeh Safari[1,2]*, George Verghese[1,2], Vito R. Ciancia[3], Daniel K. Sodickson[1,2], Seena Dehkharghani[1,2,4]**, Leeor Alon[1,2]**

[1]Center for Advanced Imaging Innovation and Research (CAI$^2$R), New York University School of Medicine, New York, NY, United States, [2]Center for Biomedical Imaging, New York University School of Medicine, New York, NY, United States. [3]LaGuardia Studios, New York University, New York, NY. [4]Department of Neurology, New York University Langone Medical Center

* **Equal contribution**

** **Corresponding authors**

Drs. Dehkharghani and Alon share co-last author and co-corresponding author responsibilities for the provided work.

Seena Dehkharghani, M.D.
Center for Biomedical Imaging
Professor of Radiology and Neurology
New York University School of Medicine
New York, NY 10016, USA
Email: Seena.Dehkharghani@nyulangone.Org

Leeor Alon, Ph.D.
Center for Biomedical Imaging
Assistant Professor of Radiology
New York University School of Medicine
New York, NY 10016, USA
Email: Leeor.Alon@nyulangone.Org

\* Corresponding, co-last authors (Emails: leeor.alon@nyulangone.org; seena.dehkharghani@nyulangone.org)

**Abstract**

Stroke is a leading cause of mortality and disability. Emergent diagnosis and intervention are critical, and predicated upon initial brain imaging; however, existing clinical imaging modalities are generally costly, immobile, and demand highly specialized operation and interpretation. Low-energy microwaves have been explored as low-cost, small form factor, fast, and safe probes of tissue dielectric properties, with both imaging and diagnostic potential. Nevertheless, challenges inherent to microwave reconstruction have impeded progress, hence microwave imaging (MWI) remains an elusive scientific aim. Herein, we introduce a dedicated experimental framework comprising a robotic navigation system to translate blood-mimicking phantoms within an anatomically realistic human head model. An 8-element ultra-wideband (UWB) array of modified antipodal Vivaldi antennas was developed and driven by a two-port vector network analyzer spanning 0.6-9.0 GHz at an operating power of 1 mw. Complex scattering parameters were measured, and dielectric signatures of hemorrhage were learned using a dedicated deep neural network for prediction of hemorrhage classes and localization. An overall sensitivity and specificity for detection >0.99 was observed, with Rayliegh mean localization error of 1.65 mm. The study establishes the feasibility of a robust experimental model and deep learning solution for UWB microwave stroke detection.

**Introduction**

Despite recent progress, stroke remains a leading cause of death and disability, disproportionately affecting low-income countries and the economically disadvantaged[1,2]. In the US alone, ~795,000 individuals are affected per annum, with 1 in every 4 people suffering a stroke in their lifetime[3]. It is estimated that ~1.9 million neurons are lost each minute during a stroke[4]; however, intravenous tissue plasminogen activator—the only thrombolytic therapy approved by the U.S. Food and Drug Administration—is successfully administered in only ~10% of patients due to strict guidelines prohibiting its use beyond 4.5 hours or in patients with hemorrhage. The resulting economic burden of stroke is thus staggering, with 34% of global total healthcare expenditure attributed to stroke, and greater than USD 56 billion in associated costs in the U.S. alone. Urgent diagnosis and intervention are therefore central pillars of contemporary management guidelines, with initial triage of stroke patients hinging first upon exclusion of brain hemorrhage by advanced neuroimaging[5–8].

Existing imaging solutions comprise primarily magnetic resonance imaging (MRI)[8,9] and computed tomography (CT)[10–12] which, while well-established, present numerous drawbacks including: 1) lack of portability for deployment to preclinical settings, engendering critical treatment delays until hospital arrival in most cases[13,14]; 2) high costs, exceeding millions of dollars for contemporary clinical CT and MR systems, proving prohibitive in many environments and perpetuating immense imbalances in global health[15]. Specifically, with two-thirds of the global population lacking accessible medical imaging, RAD-AID estimates that 3-4 billion excess deaths could be averted through improved access[16]; 3) lengthy scan and protocol durations imparting additional delays that further reduce the likelihood of favorable clinical outcomes[4,17–19] ;4) requisite levels of specialization necessary for operation and interpretation of such examinations[20,21]; and, 5) safety concerns relating to ionizing radiation in CT and strong magnetic fields in MRI[18]. While the recent introduction of CT scanners in specialized mobile stroke ambulance units[22] may alleviate some delays, their operational costs have remained untenably high for all but a small number of urban centers worldwide[23–25] . The demand for cost-effective, fast, safe, and deployable small-form diagnostic instruments is thus unmet, motivating new directions using novel sensors.

The use of low-energy microwaves as a probe of tissue dielectric properties has been explored and represents an intriguing means of characterizing tissue properties[26–29]. Several potential advantages over MRI and CT have motivated the development of such microwave imaging (MWI) systems, including: 1. extremely low operational power and lack of ionizing radiation, enabling biologically harmless use unhindered by indwelling implants and ferromagnetic objects; 2. remarkably high acquisition speeds; 3. low production cost; and 4. small form factor[30–35].

The characterization of biologic tissues using microwave radiation entails the collection of scattering (S)-parameters, based in transmission and reflection measurements, commonly achieved using vector network analyzers (VNA)[36] to drive radiation of a high directivity antenna, or with software defined radio (SDR)[37]. Most existing MWI systems obtain measurements in the low (e.g. 0.5-2.0) GHz range, owing in part to the relative ease of corresponding antenna design and fabrication[28]. With early studies casting doubt upon any additional benefits from higher *forbidden* frequencies [27] (due to exponential penetration losses from electromagnetic skin effects) as well as a generalized reduction in dielectric contrast beyond 5 GHz, low frequency MWI has remained the favored strategy[38]. Nevertheless, rigorous investigation into ultra-wideband (UWB) systems is lacking, and the potential benefits afforded by their greater resolving ability and sensitivity to superficial processes remains unknown[39,40]. The emergence of contemporary machine learning algorithms for reconstruction has further advanced the potential of microwave systems, such as for classification and localization for stroke[26,28,29,41,42]; however, it bears emphasis that past approaches to hemorrhage detection have generally been limited to models of intraparenchymal hematomas (IPH), a subclass of well-circumscribed bleeding mostly confined to the parenchymal substrate of the brain; meanwhile, more complex models recapitulating catastrophic, superficial subarachnoid hemorrhage (SAH) and other diffuse extracerebral bleeding has lagged [26,28,29,41,42],pointing to vast potential applications for systems leveraging high frequency capabilities.

We have previously shown the potential for such UWB (0.5-6.0 GHz) systems through in silico simulation and using a deep neural network for classification and localization of intracranial hemorrhage, exhibiting excellent classification (AUC 0.996) and localization (sub-millimeter error) accuracy across varied anatomical human head geometries and noise conditions[29]. However, translation from in silico environments to fully realized, physical UWB systems demands non-trivial hardware and software solutions dedicated to the specific experimental objective. In this study, we describe the conception, design, and development of a system for UWB microwave experimentation, benchtop hypothesis testing, analysis, and hardware prototyping for prediction of brain hemorrhage. A custom UWB system was engineered for operation up to 9.0 GHz and paired with an anatomically realistic human head model to evaluate morphologically variable hemorrhages under remote robotic control. The system was used to train a dedicated artificial neural net, hypothesizing robust and frequency-dependent hemorrhage localization and classification accuracy benefitting from UWB interrogation.

**Methods**

*Antenna design*

UWB microwave transmission and reception in the near field was accomplished with custom, circularly-loaded antipodal Vivaldi antennas modified from Siddiqui J, et al, having shown desirable directive properties for this antenna structure[43]. Initial simulations were conducted using the HFSS simulation environment (Ansys Inc., Canonsburg, PA, USA), where the conductor dimensions were modified to support efficient operation (S11 <-10 dB) at frequencies between 0.75 and 10 GHz. Upon arriving at the desired parameters, the antenna dimensions were 125.3 mm x 85.4 mm x 1.524 mm (length x width x thickness). The antenna was fabricated on a Rogers RO4003C substrate with a dielectric constant of 3.55 and an SMA connector was soldered to the antenna structure. Overall, a total of eight antennas were positioned in a ring, encircling a head phantom in a common XY-plane as illustrated in Figure 1a.

*Phantom design and general setup*

An anatomically realistic human brain phantom was designed, and 3D printed, based on the SAM head phantom[44]. The head model comprised a superficial (radially) outer compartment mimicking muscle and an inner compartment of *average* brain tissue, with outer dimensions of 258 mm (height) x 175 mm (width) x 122 mm (depth) and an inner compartment with dimensions of 199 mm x 145 mm x 92 mm (see SM Figure 1, *Supplemental Results*). The phantom was printed using a stereolithography 3D printer, (Phrozen Mega Sonic 8K, Phrozen Tech Co Ltd., Taiwan) leaving a aperture at its crown such that blood phantoms of varying size and morphology could be introduced to the inner compartment and navigated remotely under robotic guidance as detailed below.

Five different hemorrhage models, each simulating the dielectric properties of intracranial hemorrhage, were designed in order to introduce dielectric disturbances within the head, including: i. three distinct spherical models with diameters of 10 mm, 20 mm, and 30 mm (with inner volumes of 0.52 ml, 4.18 ml, and 14.14 ml, respectively); ii. A star-shaped model; (3.7 ml); and iii. a plus-shaped model (9.5 ml), each used to introduce complex dielectric disturbances to the head phantom during dedicated experimentation. The blood phantoms were designed and fabricated using the Formlabs Form 3B (Formlabs Inc., Massachusetts, USA) SLA 3D printer using the transparent biomed clear resin. Further details on the composition of the liquids can be found in the *Dielectric Liquid Preparation* section within the *Supplemental Materials*. The head phantom, blood phantom and positioning table are illustrated in Figure 1a and b.

*Mechanical and Robotic localization system*

To ensure fixation of the antennas relative to the head phantom, a 3D printed positioning table was designed with holders in order to immobilize the antenna array. A system of self-locking tiles was 3D printed and interlocked to meet the specifications of the phantom but readily modifiable to accommodate a wide range of experimental layouts. On the reverse face of each tile, recesses were engineered to house nuts and to provide a robust base for secure bolt fixation. Custom-designed tile holders were produced for the head phantom and antennas, and the overall configuration was designed in order to maximize stability of the phantom and antenna array during experimentation.

To ensure precise navigation, a Niryo Ned (Niryo robotique industrielle, Wambrechies, France) robotic system was used, providing a repeatability of 0.5 mm[45]. The system was equipped with self-reporting feedback on the spatial coordinates of the hemorrhage model in real time. In order to allow navigation of the phantom within the head model under robotic guidance, a firm linear rod of length 650 mm was 3D printed, articulating the hemorrhage to the robot. In order to mitigate coupling between the robot and the antenna array, the robot was elevated and placed on a grounded copper sheet with an encircling copper mesh around the opening through which the rod entered. Custom software was developed in Python for design of the automation protocol and execution of the trajectory (Figure 1a).

*Acquisition procedure*

A 2-port vector network analyzer (VNA) system (ZNB-20, Rohde & Schwarz GmbH, Munich, Germany) was used to support UWB measurements and was coupled to a switch matrix (ZN-Z84, Rohde & Schwarz GmbH, Munich, Germany) to expand the number of measurable ports to eight channels. This arrangement allowed computation of a full 8-port S-parameter network, allowing comprehensive characterization of transmission and reflection properties. The range of measurement spanned from 0.6-9.0 GHz, with a total of 8412 frequencies in *stepped* sweep mode, with 1 MHz steps. While the antenna had a reflection coefficient of -10 dB at 0.75 GHz, a lower minimum frequency of 0.6 GHz was chosen as the lower bound of the measurement sweep in order to assess for potentially valuable information content from the lower frequencies (despite the declining antenna efficiency at those frequencies, where reflection coefficients were <~-3 dB). An n-port *Unknown through - Open - Short - Match* (UOSM) calibration[36,46–48] was performed at the ends of the 8 coaxial cables to improve data quality and measurement accuracy[48].

For each acquisition, the robotic navigation system was used to translate the position of the hemorrhage phantom within the inner brain compartment of the phantom. After each translation, S-parameter measurements were conducted using the VNA and switch matrix to toggle through the antenna array before advancing the hemorrhage to its subsequent location. Importantly, when utilizing the robot in conjunction with the rod, boundaries arising from the inner wall of the head impeded the extent of translation in some dimensions, dependent upon the geometry of each hemorrhage phantom; consequently, the trajectory of the phantoms was prescribed specifically such that the peripheral-most station was made uniform and could accommodate every one of the hemorrhage phantoms without perturbing the head. We established a grid with intervals of 10 mm in the X-, Y-, and Z-directions; bounding box dimensions for the largest hemorrhage were 90 mm, 95 mm, and 60 mm, respectively. This resulted in a total of 127 positions within the head phantom, each of which was used in the random navigation trajectory of the robot-rod-phantom assembly. This procedure was repeated for three separate conditions of brain-, air- and water-filled head phantom substrate (see *Neural Network Architecture* below). Software was developed in Python to interface with the VNA and robot using vendor-provided libraries in order to automate data acquisition. The complex-valued S-parameters were processed using the open-source scikit-rf library[49]. A front-end interface was developed using PySide6 libraries providing real-time plotting of the scattering measurements, robot movements, and tracking of captured data. A desktop PC was used to interface to the VNA using a USB port and to the robot over Ethernet cable. All acquisitions were performed with the head phantom filled with human brain-mimicking (interior compartment) and muscle-mimicking (exterior compartment) liquids for the five different hemorrhage phantoms. The acquisition time for each point when using a 100 kHz IF bandwidth was 72 seconds, with a 12-second delay (measured empirically) imposed after each translation to mitigate motion-related noise.

*Neural network architecture*

The neural network architecture proposed in this study was tailored to interpret complex S-parameter data. The architecture was segmented into two cardinal components: the convolutional layers and the terminal fully connected segments (Figure 2). The former is characterized by a Residual ConvBot (Figure 2d) and a series of bottleneck residual blocks[50] (Figure 2c), termed "ResidualBlocks". These ResidualBlocks function as feature extractors, converting intricate, high-dimensional data into lower-dimensional form while retaining important information. The unique arrangement of these ResidualBlocks ensured a balance between model intricacy and operational output. The terminal components of the network were anchored by a series of fully connected layers. A linear layer served as an integration point for the output matrix of the ConvBot (Figure 2c). The layers were designed for discrete predictive functions, including pinpointing the location coordinates of the hemorrhage and classifying the hemorrhage shape. Full details of the network architecture can be found in the *Neural Network Architecture* section in the *Supplemental Materials*.

*Neural network training and validation*

The network included three outputs comprising two classifications (multi-class classification with SoftMax) and a three-dimensional regression. Multi-task learning network training was performed with the Adam optimizer[51] using the following loss function:

$$loss = \frac{(1-\alpha)}{N} \sum (L_1 + MSE + dist(p,\hat{p})) + CE_1 + CE_2,$$

$$\alpha = \begin{cases} 0 & hemorrhage \\ 1 & no\ hemorrhage \end{cases}$$

where $p$ is the ground truth location and $\hat{p}$ is the predicted location. $L_1, MSE$, and $dist$ are the 1-norm, mean squared error, and distance difference between ground truth and predicted locations, respectively. $N$ denotes the batch number, $CE$ denotes the cross-entropy loss, and $\alpha$ ensures that the gradients for localization become zero in those cases for which the hemorrhage did not exist (i.e., *no-hemorrhage* class), while still allowing training for the classification task. The initialization of the learning rate of the Adam optimizer was set to 0.0001, which was decreased every hundred epochs by a factor of $\frac{1}{10}$. The network was trained on all media for 500 epochs with batch sizes of 48 over 70% of the data and tested on the remaining 30% of data for the brain-filled head phantom only (i.e., air- and water-filled phantom datapoints were used as an augmentation to reduce potential overfitting). From the training set, 10% was held out for validation during training. The hold-out method was used for data separation.

*Evaluation of frequency range and antenna number*

An ablation study was conducted to investigate the effects of varying the frequency range and number of antennas on network performance. Consistency in all other hyperparameters was maintained to isolate the impact of the selected variables. For the frequency range evaluation, four distinct spans were chosen: 0.6-1.5 GHz, 0.6-3 GHz, 0.6-6 GHz, and the full range of 0.6-9 GHz. For each span, we adapted the S-parameter matrix to correspond to the chosen frequency band by retaining the pertinent rows and columns and setting others to zero before network training commenced. After zeroing out the undesired frequencies, classification and localization errors were evaluated. For the evaluation of the number of antennas used for network performance, the following combinations of antennas were tested: i. antennas 1 and 3; ii. antennas 1, 3, and 5; iii. antennas 0, 1, 4, and 5; and iv. all 8 antennas in the array. As with the exploration of frequency dependency, the S-parameter matrix was modified for each configuration, with relevant rows and columns preserved and non-pertinent ones nullified before network training.

*Statistical analysis*

Network performance was summarized with a confusion matrix for multi-class classification using frequency spans of 0.6-1.5 GHz, 0.6-3.0 GHz, 0.6-6.0 GHz, and 0.6-9.0 GHz. The specificity and sensitivity of hemorrhage detection was computed. To assess the fidelity of the localization predictions, the probability distribution function (PDF) of the error in X- and Y-directions, XY-plane, and XYZ-space were computed and plotted for frequency spans of 0.6-1.5 GHz, 0.6-3 GHz, 0.6-6 GHz, and 0.6-9 GHz. Distance error in hemorrhage localization was expressed in a normalized histogram of the distance error, with the single variate (X- and Y-directions) data fitted to a folded normal distribution function[52], and the bi-variate (in XY-plane) distance errors fitted to a Rayleigh distribution function[53]. Notably, because an array of distributed channels did not exist in the z-direction, a declining accuracy was anticipated in z-direction, which was confirmed in preliminary analysis. All statistical analysis was performed using the SciPy[54] library in Python.

**Results**

Excellent performance was demonstrated using the complete antenna array together with the entire range of UWB frequencies. Specifically, sensitivity and specificity of >0.99 were observed for hemorrhage detection, with an overall median localization error of 1.67 mm. Full results for classification and localization tasks are summarized in Figure 3. Specifically, the larger frequency spans, of 0.6-6.0 and 0.6-9.0 GHz, outperformed the narrower ranges in classification, Notably, despite excellent overall performance in the multi-class classification task for the 0.6-6.0 GHz range, there were a small number of instances for which a plus-shaped phantom was misclassified as a 20 mm diameter sphere, while the full frequency range exhibited better performance in discriminating spherical and non-spherical phantoms. Together, the results suggest that the extended UWB frequency ranges generally provide a superior combination of localization (again, demonstrating a median distance error of 1.67 mm in the XY-plane, Table 1) and classification performance (Figure 3).

Among the individual sub-ranges of frequencies, the 0.6-6.0 GHz span yielded the best overall performance in classification of hemorrhage classes, peaking in its identification of the smallest (10 mm diameter) spherical phantom at a classification accuracy of 98.25% and a misclassification rate of only 1.75%. Overall multi-class classification for this frequency span was 94.88%. The lowest individual performance was 81.87% for the frequency span 0.6-3.0 GHz. Importantly, there was high overall sensitivity to the morphological characteristics of the hemorrhage phantoms; specifically, using the full frequency rage (0.6-9.0 GHz) there were no instances of a spherical phantom of any size being misclassified as a non-spherical phantom, or vice-versa. Further, a strong capacity for discrimination of varying spherical phantom volumes was observed, with no cases of the smallest 10 mm sphere being misclassified as the largest 30 mm sphere, or vice-versa. Only a small percentage (10.53%) of hemorrhages in the 20 mm class were misclassified as either a 10 mm or 30 mm hemorrhage. The single worst performing classification task was observed in prediction of the star-shaped phantom using the 0.6-3.0 GHz sub-range, for which 33.33% of cases were misclassified as plus-shaped phantoms; however, only 6.14% of such cases were misclassified using the frequency sub-range of 0.6-6 GHz (Figure 3a).

For the star-shaped phantom, although the volume was 3.7 ml (volumetrically almost identical to the 3.5 ml of the middle-sized sphere), excellent discrimination (>99.9%) was present between the two, further suggesting sensitivity to morphologic characteristics irrespective of volumes. The largest class-wise false-positive and false-negative rates between the star-shaped phantom and the plus-shaped phantom were 9.65% and 6.14% (star-shaped predictions), respectively, for the best performing network. The histogram of the localization error in XYZ- and XY-space is illustrated in Figure 3c, demonstrating a median error of 5.68 mm and 1.90 mm in the XYZ- and XY-space. As anticipated from our antenna array's geometry, the histogram for the XY distance error reveals a reduced error compared to the XYZ error distribution, related to the arrangement of the antennas in the XY plane around the head phantom, rather than stacked in the Z-direction.

The error distribution for frequency spans 0.6-1.5 GHz, 0.6-3 GHz, 0.6-6 GHz, and 0.6-9 GHz in the X-direction, Y-direction, and XY-plane are shown in Figure 4. The PDF of the error in the X- and Y-directions and in the XY-plane both revealed squared residuals of less than 0.24. For the narrowest frequency span of 0.6-1.5 GHz, the distribution was the widest, achieving a maximum folded normal mean error of 2.98 mm. Conversely, the narrowest distribution (mean of 1.14 mm was obtained for the widest frequency span of 0.6-9.0 GHz. Figure 4b depicts the error distribution for the Y axis, showing that the best performance was observed for the network trained on the widest range (1.05 mm), while the least favorable performance was observed on the narrowest range (2.01 mm). Figure 4c illustrates the distance error in the XY plane; here, a similar trend was present, with the strongest performance associated with the widest frequency range of 0.6-9.0 GHz (1.65 mm). A detailed summary of the median error across frequency spans can be found in Table 1.

The relationship between the number of antennas utilized and the stroke localization error in the X-direction, Y-direction, and XY-plane is illustrated in Figure 5. Stroke localization error was analyzed for the full frequency span of 0.6-9.0 GHz, demonstrating a reduction of localization error as the number of antenna elements was increased. In the case of two antennas 90-degree apart (antenna 1 and 3), median localization error was 2.62 mm in the XY-direction; however, incrementally, as the number of antenna elements was increased to 8-elements, the error was reduced to 1.67 mm. With regards to the X-, Y-, and XY-directions, an increase in the error in each individual component was observed. A summary of the median error in for different antenna combinations is summarized in Table 2.

**Discussion**

The measurement of tissue electrical properties has long been a focus of interest for the biomedical community, motivated by potential applications in a multitude of neurological diseases. While past studies have reported the potential of microwave-based stroke characterization[26,28,29,41,42], this represents, to our knowledge, the first study leveraging such operating bandwidths (~8.4 GHz), and the only study to date utilizing a dynamic hemorrhage model (comprising five different shapes and 127 positions).

Our system was constructed in the context of near-field stroke classification and localization, with the goal of utilizing supervised deep learning approaches to learn dielectric signatures. Because supervised learning was used to classify and localize hemorrhage models, the experimentation necessitated a flexible system capable of being navigated throughout the head, such that many data points could be acquired. This system included the fabrication of UWB antennas and creation of a mechanical system to mobilize blood-mimicking phantoms of varying shapes and sizes inside a multi-compartment head model. These were developed to assess and improve UWB antenna arrays, as well as develop methods capable of learning signatures of dielectric changes in tissue. In previous in-silico reports, we have shown the potential advantages of UWB approaches for the detection of dielectric disturbances in head models[29], and in this work, a physical realization of such a system was constructed.

A residual convolutional network structure was used as the core for higher dimensional feature extraction, and for classification and localization, a collection of fully connected layers was used. While there exist many potential approaches for learning from the complex-valued S-parameter space, our architecture was conceived as a small working model with only 1.7 million parameters. In examining the classification performance more closely, a sensitivity and specificity of 1.0 was achieved for hemorrhage detection (Figure 3b). Further, when utilizing UWB, none of the spheres, irrespective of their size, were mistakenly predicted as star- or plus-shaped phantoms (Figure 3a, 0.6 - 9.0 GHz), pointing to a remarkable sensitivity for complex morphological characteristics, even in excess of volumes alone, which were nearly identical for the disparately shaped star phantom (3.7 ml) and middle-sized sphere (3.5 ml). Similarly, predictions for the star-shaped phantom were only misclassified in any instance as a plus-shaped phantom but never as a spherical hemorrhage. The data supports that wider frequency ranges in many cases improved geometric discrimination of hemorrhages, with the most accurate classification in the 0.6 - 6.0 GHz sub-range.

The system yielded a median localization error for hemorrhage of 5.68 mm and 1.9 mm in the XYZ-space and XY-plane, respectively. This difference in error between the XYZ-space and XY-plane was not unexpected in view of the limited number of antennas in the Z-direction, thus limiting information in that direction. Where localization errors did occur, the effects are traceable, at least in part, to errors of the robotic system itself. Specifically, the mean XY positioning error of the robot with the attached rod was ~1.5 mm at baseline, which places in more remarkable perspective the localization accuracy of the full model. It should be noted that the best performing frequency span for the localization task was using 0.6-9.0 GHz and perfect performance in discriminating spherical and non-spherical phantoms were achieved. The results suggest that the best choice of UWB frequency ranges is likely dependent on the network architecture and the task in question. Notably, while past studies have suggested the presence of *forbidden frequencies* between 1.5 and 4 GHz[27], our results suggest the presence of valuable information content encoded with these frequencies and unlocked by the neural network in use in our study.

With regards to the phantom liquids used for the brain, muscle, and blood phantoms, recipes for those liquids were chosen such that the dielectric properties match the dielectric properties of their respective biological tissues at 2.45 GHz (supplemental materials, Dielectric liquid preparation section). This single-frequency selection was chosen due to the fact that measurement of dielectric properties of materials is more challenging at high frequencies (> 6 GHz), necessitating specialized measurement equipment. Second, to the best of our knowledge, it is extremely challenging to match the conductivity and permittivity of tissues using liquids at UWB frequencies. Therefore, we matched the dielectric property of tissues [55] at a single frequency and within an error margin of 10%. Future investigations, of course, can include more elaborate phantom designs that match a wider range of frequencies using more advanced recipes.

Overall, the performance of our system benefitted from several unique attributes, including the successful development of UWB capabilities spanning 0.6-9.0 GHz. The results document, for the first time, the benefits of localization accuracy enabled through the use of realized UWB hardware alongside deep neural network, as well as hemorrhage classification with dynamic phantoms. The apparent sensitivity to morphologic characteristics, in excess of simple dimensional or volumetric features, points strongly to the benefits of deep learning tools in the exploration of UWB microwave scattering.

**Acknowledgments**

This work was supported in part by NIH grant P41-EB017183.

**Figures**

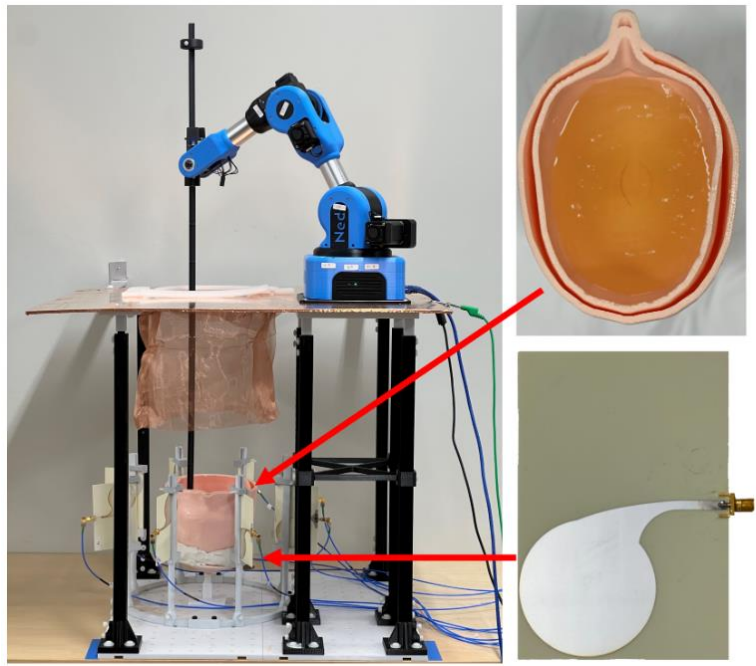

(a)

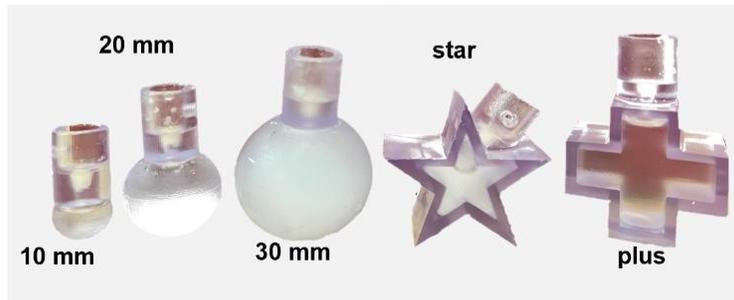

(b)

**Figure 1.** (a) robotic system used for phantom navigation within the head model. A 65 mm rod was used to connect the robot to the blood phantom (left). A 3D printed head model was used and filled with brain-mimicking dielectric liquid (top-right, with its position in the model shown by paired red arrows) and encircled by an array of custom UWB antipodal-Vivaldi antennas for near-field measurements (bottom-right). (b) 3D printed blood phantoms including spherical hemorrhage models of varying inner diameter, as well as morphologically more complex star- and plus-shaped models.

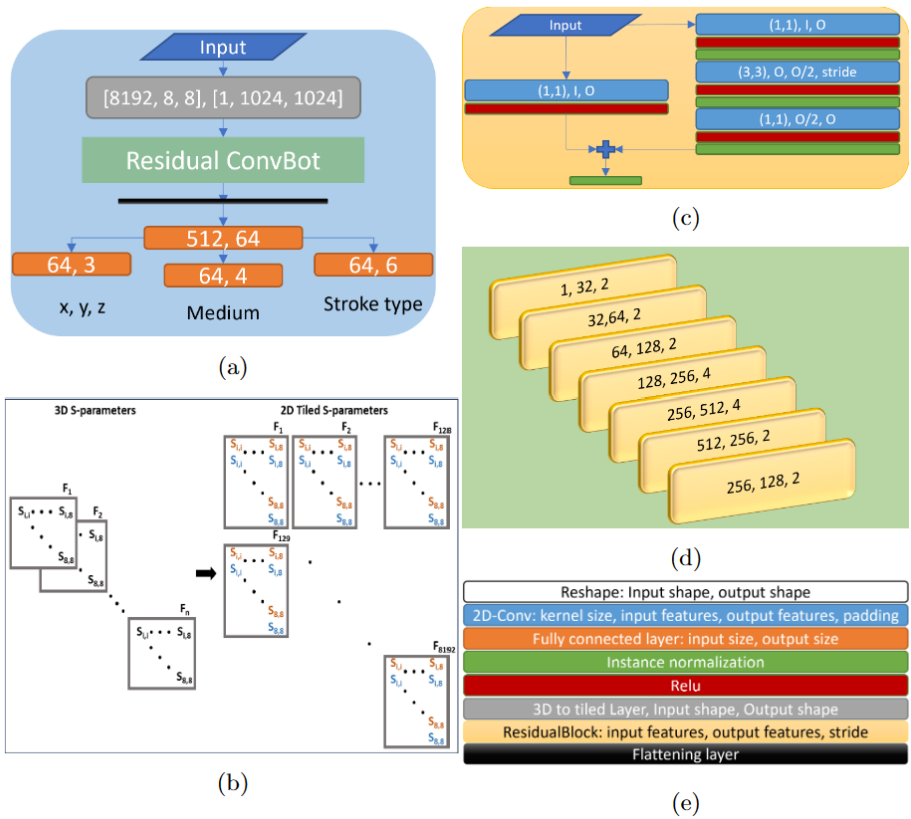

**Figure 2.** Network architecture. (a) The RobotNet structure with (b) a tiling layer for preprocessing of data. (c) Seven Residual Blocks were used as building blocks for the Residual ConvBot (d) design. The architecture contains several fully connected layers for the final output prediction. (e) dictionary illustrating the layer structure.

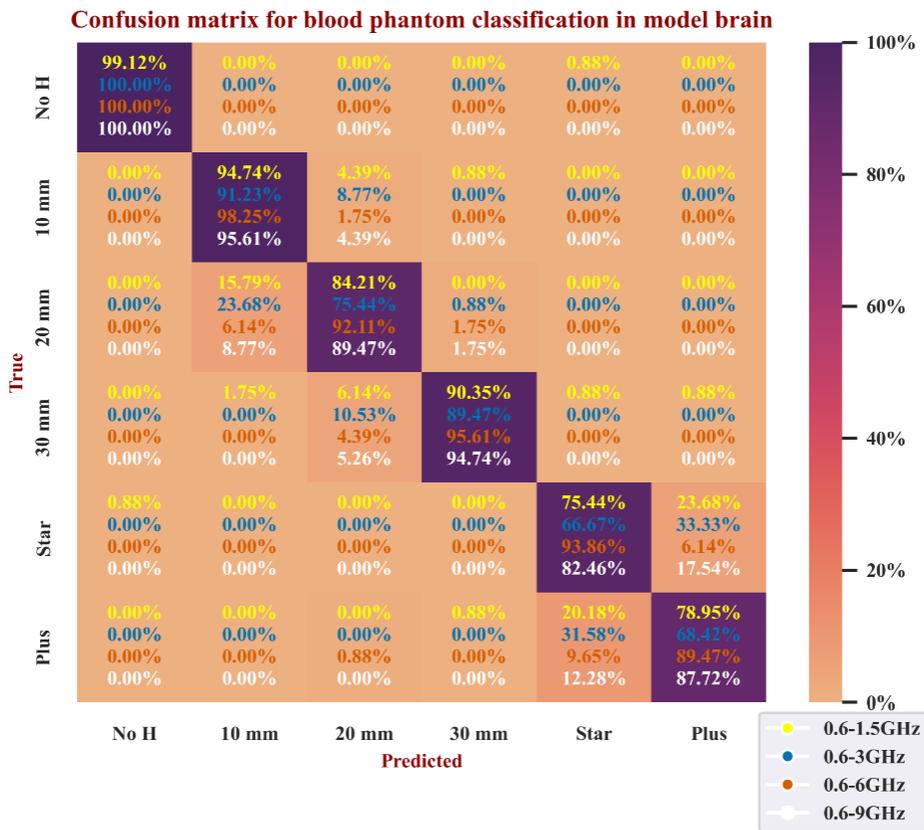

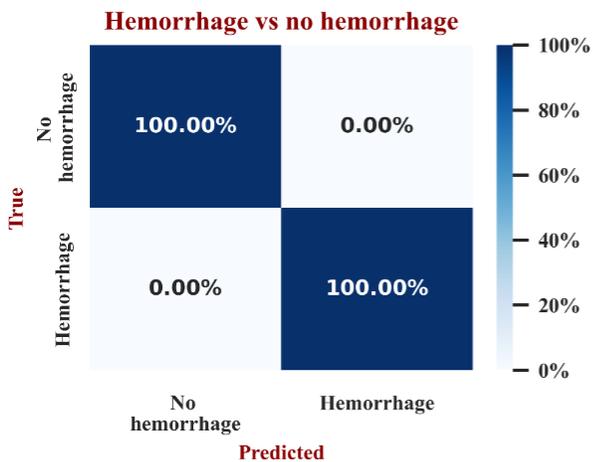

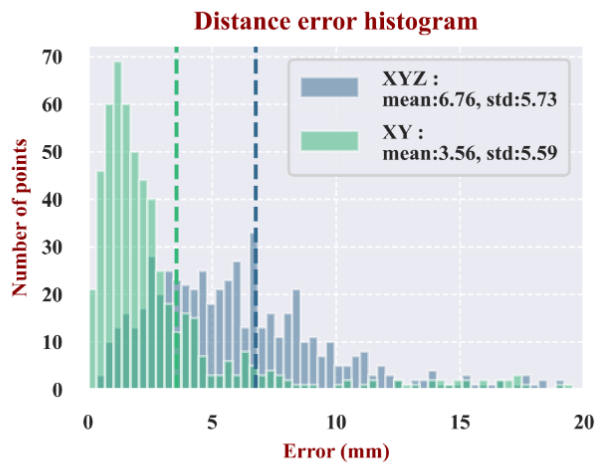

**Figure 3.** (a) Classification confusion matrix for varying blood phantoms as well as the case of *no-hemorrhage* for network training using different frequency spans. (b) Overall classification performance for hemorrhage (all phantoms combined) versus no hemorrhage, demonstrating overall sensitivity and specificity of >0.99. (c) histogram of the localization error in the XYZ-space and XY-plane.

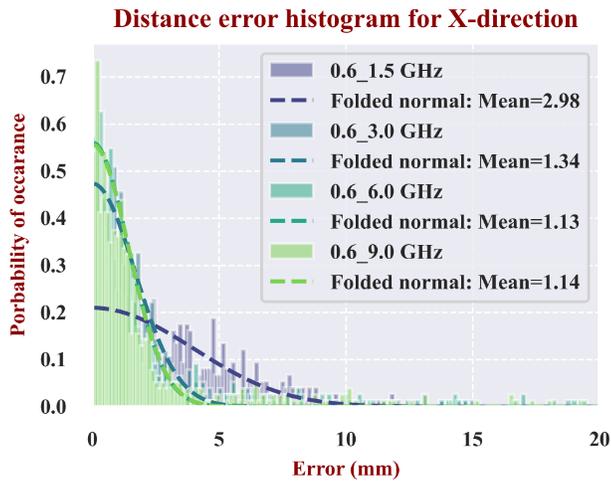
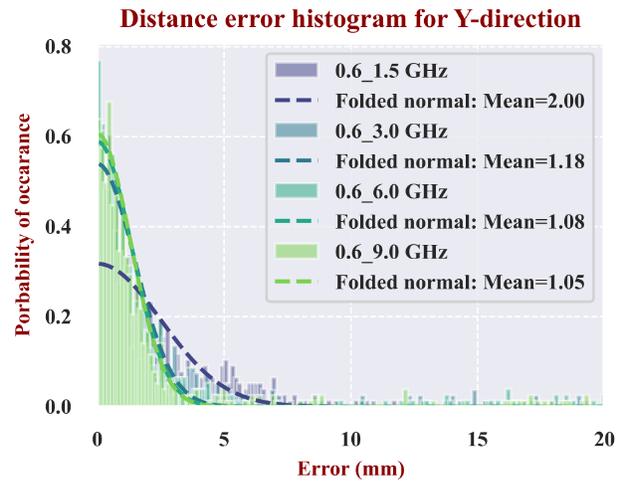
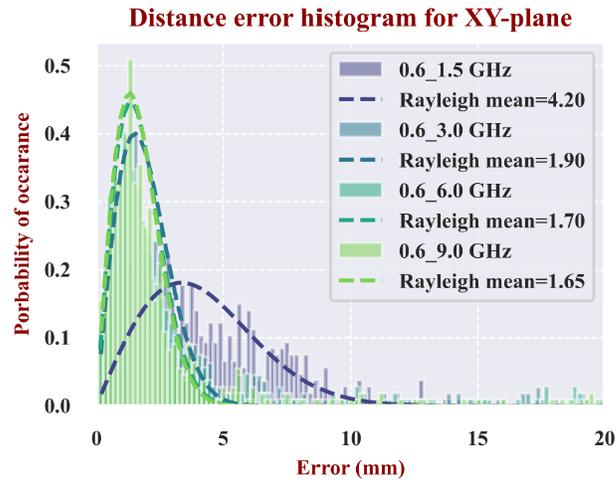

**Figure 4.** PDF of the distance error histogram for the X-direction (a), Y-direction (b), and XY-plane (c). The mean fitted Folded Normal distributions for X and Y are shown to decrease as the frequency span increases.

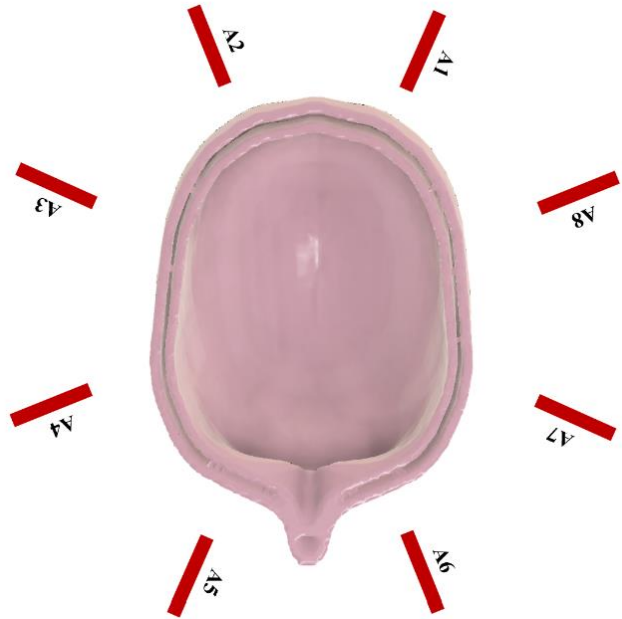

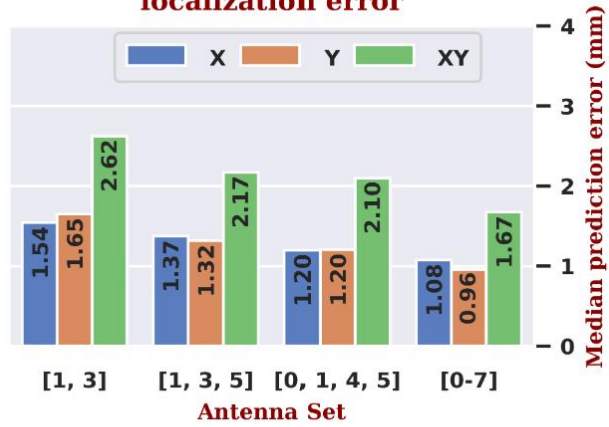

**Figure 5.** (a) Placements of the antennas relative to the head phantom. (b) Model performance in relation to the number of antennas employed in the experiment, demonstrating the effect of increasing antenna number on localization accuracy.

| Frequency span (GHz) | Median error (mm) | | | | |
|---|---|---|---|---|---|
| | X | Y | Z | XY | XYZ |
| 0.6 – 1.5 | 2.52 | 1.67 | 3.93 | 3.82 | 6.61 |
| 0.6 – 3 | 1.13 | 1.04 | 4.46 | 1.9 | 5.68 |
| 0.6 – 6 | 1.03 | 0.99 | 5.66 | 1.69 | 6.79 |
| 0.6 – 9 | 1.08 | 0.96 | 5.54 | 1.67 | 6.85 |
| 1.5 – 3 | 1.17 | 1.17 | 5.02 | 1.99 | 6.44 |
| 3 – 6 | 1.19 | 1.22 | 6.47 | 2.02 | 7.82 |
| 3 – 9 | 1.35 | 1.32 | 6.17 | 2.37 | 7.33 |
| 6 – 9 | 2.38 | 2.05 | 8.89 | 3.96 | 10.95 |

**Table 1.** Median Error for different frequency spans.

| Antennas in array | Median error (mm) | | | | |
|---|---|---|---|---|---|
| | X | Y | Z | XY | XYZ |
| [1, 3] | 1.54 | 1.65 | 7.47 | 2.62 | 9.69 |
| [1, 3, 5] | 1.37 | 1.32 | 4.85 | 2.17 | 6.37 |
| [0, 1, 4, 5] | 1.20 | 1.20 | 5.62 | 2.10 | 6.96 |
| [0 – 7] | 1.08 | 0.96 | 5.54 | 1.67 | 6.85 |

**Table 2.** Median error for different antenna numbers in the array.

# Supplemental materials: An experimental system for detection and localization of hemorrhage using ultra-wideband microwaves with deep learning

Dielectric liquid preparation: liquids used inside the phantoms were prepared to match the dielectric properties of tissues at 2.45 GHz (the upper limits of our VNA with the dielectric probing system (Keysight 85070E dielectric probe kit) with the slim probe) designed for liquid materials. This frequency was selected as matching the exact properties of tissues across a UWB frequency range is not currently possible. The dielectric properties were targeted according to Gabriel et al. [55] and are summarized in SM Table 1, where brain properties were the average of gray and white matter.

| Tissue | $\sigma$ (s/m) Theoretical | $\sigma$ (s/m) Measured | $\varepsilon_r$ Theoretical | $\varepsilon_r$ Measured |
|---|---|---|---|---|
| Brain | 1.5 | 1.5 | 42.6 | 40.3 |
| Muscle | 1.7 | 1.6 | 52.7 | 56.6 |
| Blood | 2.5 | 2.3 | 58.3 | 60.2 |

SM Table 1. Theoretical and measured dielectric properties used inside the modified SAM phantom.

Dielectric values were achieved by mixing the ingredients shown in SM Table 2.

| Tissue | Water (%) | Tween 20 (%) | NaCl (%) |
|---|---|---|---|
| Brain | 55 | 45 | – |
| Muscle | 77 | 22.8 | 0.2 |
| Blood | 81 | 18.5 | 0.5 |

SM Table 2. Ingredients, with their respective percent in weight, used in the brain, muscle and blood liquids.

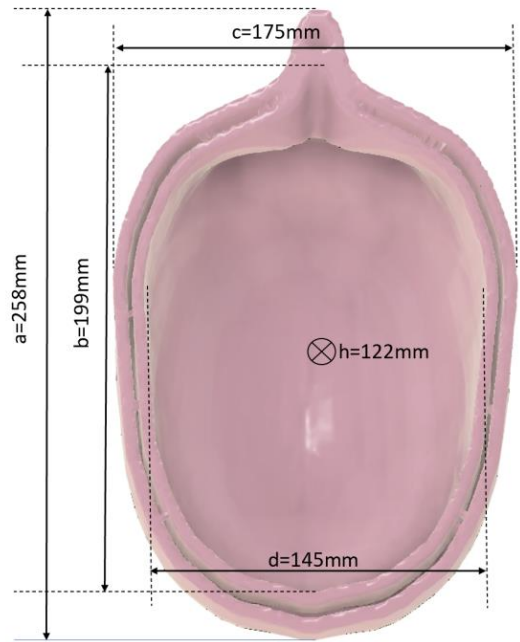

Figure SM 1. Modified, multi-compartment Head Phantom. The outer dimensions are represented by 'a' (length) and 'c' (width), while the inner dimensions are denoted by 'b' (length) and 'd' (width). The depth to the deepest point is specified by 'h'.

Neural network Architecture

The neural network deployed in this study comprised distinct blocks and layers specifically tailored to interpret and process complex number inputs. Delving deeper into the architecture of RobotNet (Figure 2a), three core sections were present:

1. This section began with an input layer which took in a complex matrix sized [8192, 8, 8] and reshaped it into a non-complex matrix sized [1, 1024,1024]. As illustrated in Figure 2b, this transformation ensured that every [8x8] segment in the new format was adjacently positioned, resembling tiles. This tiling strategy retained the spatial coherence of adjacent elements in the original matrix. The matrix strategically placed the magnitude of the complex number in even rows and its phase in odd rows, facilitating the subsequent convolutional layers in discerning their mutual relations.

2. Central to this section was the "Residual ConvBot" block, a blueprint of which is presented in Figure 2d. It comprised seven sequential "ResidualBlock" units (refer to Figure 2c). The number of ResidualBlocks (seven) was decided based on the batch number and the GPU memory size. The Residual ConvBot transformed its input and condensed it into 512 discernible features, which were later fed to fully connected layers.

3. The compact feature vector from the Residual ConvBot was relayed to a linear layer, where it was whittled down from 512 input features to 64 output features. These outputs then segued into three specialized fully connected layers:

    1. The first, a regression model, predicted X-, Y-, and Z-coordinates, outputting three values.

    2. The second, a classification model, classified among three distinct mediums and an 'unknown' category, resulting in four outputs.

3. The third layer distinguished between stroke types, offering six outputs that delineated specific stroke classifications and a no-stroke category.

Zooming back into the "Residual ConvBot" (Figure 2d), it stood as an essential part of the network. Housing a series of "ResidualBlock" units, it reduced the dimensionality of its input. Beginning with an input size of [1, 1024,1024], the ConvBot altered the input dimensions through a calculated reduction—thrice halved, twice quartered, and again halved. Resultantly, the first dimension grew to 512 and was pared down to 128. By its conclusion, the ConvBot outputted a [128, 2, 2] sized matrix, which then underwent the previously detailed processing.

In essence, ConvBot was the network's premier feature extractor. It distilled high-dimensional data into a succinct space without compromising essential input information. This reduction not only mitigated off overfitting but also bolstered the network's capacity to generalize, improving its ability to handle various input scenarios.

The ResidualBlock stood as the basic block of the Residual ConvBot. Every ResidualBlock was intricately fashioned to contract dimensionality within the convolutional arena. Inspired by the refined ResNet architecture, our ResidualBlock was designed to downscale matrix dimensions while maintaining the focus on preserving both magnitude and phase within the matrix. Diving into its structure, as portrayed in Figure 2c, two branches are apparent:

1. Primary Branch:
   - Started with a 2D convolution featuring a 1x1 kernel. Here, the input feature count was seamlessly transformed to match the designated output feature count.
   - Subsequent to this, a ReLU activation function was invoked.
   - The culmination of this branch's computations served as a residual, which was later fused with the terminal layer of the secondary branch.

2. Secondary Branch:
   - Beginning with 2D convolution brandishing a 1x1 kernel, wherein the output features were systematically pared down to half of the envisioned count.
   - This was succeeded by both a ReLU activation and an Instance Normalization layer.
   - A 2D convolution with a 3x3 kernel was then employed, ensuring consistent feature counts at both input and output stages (O/2, O/2).
   - Another bout of ReLU activation and Instance Normalization ensued.
   - Of note is the fact that any dimensionality contraction (or stride) exclusively occurred at this juncture.
   - The branch's last step involved a 1x1 convolution that restored the feature count to its anticipated volume (O/2, O), seamlessly complemented by a ReLU and another Instance Normalization.

The results of the two branches are summed, leading to ResidualBlock's final output. By centralizing dimensionality contraction within the feature realm, the ResidualBlock succeeded in slashing the model's overall parameter tally. This design was chosen to simplify structure and minimize overfitting pitfalls.